\documentclass[
 reprint,amsmath,amssymb,aps,
prl,
]{revtex4-2}

\usepackage{graphicx}
\usepackage{dcolumn}
\usepackage{bm}
\usepackage{hyperref}
\hypersetup{colorlinks, linkcolor={blue}, citecolor={blue}, urlcolor={blue}}
\usepackage{xcolor}
\usepackage{pifont}

\begin{document}
\title{Multiferroic-like Quasiparticles in Ferroelectrics}

\author{Ping Tang$^1$}
\email{tang.ping.a2@tohoku.ac.jp}
\author{Gerrit E. W. Bauer$^{1,2}$}
\affiliation{$^1$WPI-AIMR, Tohoku
University, 2-1-1 Katahira, Sendai 980-8577, Japan}
\affiliation{$^2$Institute for Materials Research and CSIS, Tohoku University, 2-1-1 Katahira, Sendai 980-8577, Japan} 

\date{\today}
\begin{abstract}
Multiferroics are materials with coexisting electric and magnetic orders that are of central importance for fundamental research and technological applications. Unfortunately, intrinsic multiferroics that operate at room temperature remain rare due to an apparent incompatibility between magnetism and ferroelectricity. Here we predict that a pure ferroelectric support multiferroic-like quasiparticles, termed ``multiferrons", that simultaneously carry \emph{static} magnetic and electric dipoles. The electric dipole moment emerges from the parity-odd anharmonicity of the ferroelectric dynamics, while the magnetic moment has both paramagnetic and diamagnetic origins generated by circularly polarized transverse fluctuations of the ferroelectric polarization. In contrast to the established ``dynamical multiferroicity" of circularly polarized phonons, which involve only \emph{oscillating} electric dipoles, multiferrons cause, apart from Zeeman and Einstein-de Haas effects, a linear dc Stark response, giant electric-field-tunable second-harmonic generation in the THz-frequency regime, and a finite magnetoelectric cross coupling. Multiferrons open a new route toward nonlinear THz optical applications and offer multiferroic functionalities with simple ferroelectrics.
\end{abstract}

\maketitle
\emph{Introduction---}Multiferroic materials that combine magnetism and ferroelectricity may lead to multifunctional electronic devices such as low-power magnetoelectric memories~\cite{hur2004electric,eerenstein2006multiferroic,dong2015multiferroic,vopson2015fundamentals}, in which the cross-coupling between ferroic orders enables electric-field control of magnetization and vice versa. However, intrinsic room-temperature multiferroics remain scarce, BiFeO$_3$ being among the few known candidates~\cite{wang2003epitaxial, PhysRevB.71.014113, zhao2006electrical, catalan2009physics}. 

Conventional multiferroicity requires simultaneous breaking of time-reversal ($\mathcal{T}$) and inversion ($\mathcal{P}$) symmetries in the ground state. In contrast, a time-dependent electric polarization in \emph{nonmagnetic} materials can generate a magnetization~$\mathbf{M}\sim(\mathbf{P}\times\partial_{t}\mathbf{P})$ through a circulating displacement current~\cite{dzyaloshinskii2009intrinsic}, leading to paramagnetism in ferroelectrics~\cite{lashley2007heat}. Likewise, circularly polarized (and sometimes called chiral) phonons carry an orbital angular momentum~\cite{levine1962note,PhysRevLett.112.085503,PhysRevLett.115.115502,PhysRevLett.121.175301,zhu2018observation,tauchert2022polarized,ishito2023truly,zhang2024observation,zhang2025measurement,juraschek2025chiral,ueda2025chiral} and an associated magnetic moment~\cite{PhysRevMaterials.3.064405,PhysRevResearch.3.L022011}. These effects fall under the so-called ``dynamical multiferroicity"~\cite{PhysRevMaterials.1.014401,PhysRevLett.122.057208,basini2024terahertz,7lm1-wm3y}. Despite the small gyromagnetic ratio for ionic cyclotron motion, large phonon magnetic moments (up to the Bohr magneton $\mu_{B}$) have been reported in a broad range of systems~\cite{schaack1975magnetics,schaack1976observations,schaack1977magnetic,cheng2020large,PhysRevLett.128.075901,hernandez2023observation,wu2023fluctuations, luo2023large,PhysRevB.109.155426, lujan2024spin} and attributed to various mechanisms~\cite{PhysRevLett.127.186403,PhysRevB.107.L020406,PhysRevB.110.094401,PhysRevB.110.024423,mustafa2025origin,chen2025geometric,chaudhary2025anomalous,6m7v-p99w}, offering a promising avenue to develop the functionalities of phonon-based magnetism~\cite{davies2024phononic}.

The collective elementary excitations of ferroic orders can be described as quasi-particles that transport energy as well as order parameters~\cite{landau2013statistical}. Magnons, the quanta of spin-wave excitations, carry spin angular momentum or magnetic dipoles and have been extensively studied in spintronics~\cite{kruglyak2010magnonic,chumak2015magnon}. Their electric counterparts in ferroelectrics, termed ferrons~\cite{PhysRevB.106.L081105, bauer2022magnonics, PhysRevApplied.20.050501, PhysRevB.109.L060301}, enable electric-dipole transport~\cite{PhysRevLett.126.187603,PhysRevLett.128.047601}, electric field control of thermal currents~\cite{PhysRevB.106.L081105,wooten2023electric,zhao2025thermal}, and nonlocal drag thermoelectricity~\cite{PhysRevB.107.L121406}. Recent experiments report the observation coherent~\cite{choe2025observation,zhang2025electric} and incoherent~\cite{shen2025observation} ferrons generated by optical and electrical stimuli, respectively. While magnons and ferrons are typically regarded as distinct elementary excitations, a fundamental question arises: do multiferroic-like quasiparticles exist that combine the properties of both in \emph{non-multiferroic} materials?

\begin{figure}
    \centering
    \includegraphics[width=6 cm]{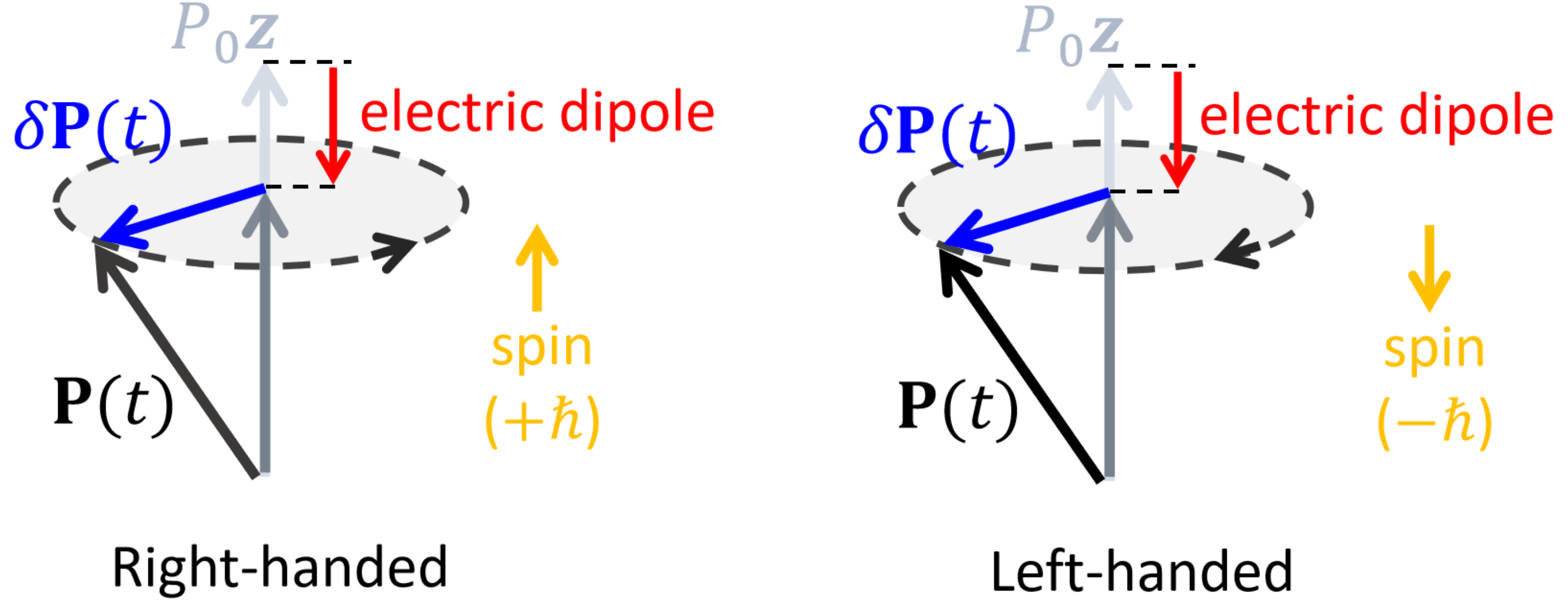}
    \caption{Schematics of two multiferron modes corresponding to right- and left-handed circulations of the transverse polarization fluctuations $\delta\mathbf{P}(t)$ around the spontaneous polarization $P_{0}\mathbf{z}$. Each mode carries a quantized angular momentum ($\pm\hbar$) as well as a \emph{net} electric dipole (red arrows) that reduces the ground state polarization.}
    \label{Fig-multiferron}
\end{figure}

In this Letter, we answer this question affirmatively by proposing a ``multiferron" chimera quasiparticle that carries both \emph{intrinsic} dc electric and magnetic dipole moments. We focus on a homogeneous ferroelectric with axial symmetry around its ground state electric polarization. On top of conventional longitudinal ferron modes, multiferrons emerge from circularly polarized transverse fluctuations of the electric polarization that acquire a net electric dipole through parity-odd anharmonicity induced by the ferroelectric order, as illustrated in Fig.~\ref{Fig-multiferron}. The resultant magnetic susceptibility contains both paramagnetic and diamagnetic contributions: the former is generated by the intrinsic angular momentum associated with the circular polarization motion, while the latter is analogous to the orbital diamagnetism of electrons. 

The proposed multiferrons differ fundamentally from electromagnons in multiferroics~\cite{pimenov2006possible,PhysRevLett.98.027202,PhysRevLett.98.027203,PhysRevLett.102.047203} and from the circularly polarized phonons that cause the dynamical multiferroicity~\cite{PhysRevMaterials.1.014401,PhysRevLett.122.057208,basini2024terahertz}. The former are electroactive magnetic excitations that require both magnetic and ferroelectric order, while the latter on time average they do not carry a net electric dipole moment. In contrast, we show that purely ferroelectric materials support multiferrons that give rise to a variety of phenomena, including the Zeeman and Stark effects, electric-field-tunable second-harmonic generation, a finite magnetoelectric coupling, and the Einstein-de Haas effect. These predictions open many experimental avenues to test our multiferron concept.


\emph{Model---}Ferroelectricity may arise, for instance, from a structural phase transition in displacive ferroelectrics through the condensation of a soft optical phonon~\cite{bilz1980microscopic} or from a rearrangement of the electronic charge distribution~\cite{PhysRevB.54.17452,ishihara2010electronic}, e.g., in the rare-earth ion oxide LuFe$_2$O$_4$~\cite{ikeda2005ferroelectricity}. Irrespective of the microscopic origin, ferroelectric materials are well described by the phenomenological Landau-Ginzburg-Devonshire (LGD) theory~\cite{chandra2007landau}, in which the free energy density $f[\mathbf{P}]$ is a functional of the electric polarization $\mathbf{P}$. We begin with the effective Lagrangian density for the polarization dynamics in the presence of external electric ($\mathbf{E}$) and magnetic ($\mathbf{B}$) fields~\cite{sivasubramanian2004physical,dzyaloshinskii2009intrinsic,riseborough2010quantum}:
\begin{equation}
\mathcal{L}=\frac{1}{2}m_{p}\left(\frac{\partial\mathbf{P}}{\partial t}\right)^2+\Lambda_{p}\mathbf{B}\cdot\left( \mathbf{P}\times\frac{\partial\mathbf{P}}{\partial t}\right)-f\left[\mathbf{P}\right]+\mathbf{E}\cdot \mathbf{P},\label{Lag}
\end{equation}%
where $m_{p}$ is the polarization inertia, and $\Lambda_{p}\mathbf{P}\times\partial_{t}\mathbf{P}$ is the magnetization generated by the circulating displacement current \(\sim\partial_{t}\mathbf{P}\), with $\gamma_p=\Lambda_{p}/m_{p}$ the associated gyromagnetic ratio~\cite{PhysRevMaterials.1.014401,PhysRevMaterials.3.064405}. Depending on the relative ionic and electronic contributions, $\gamma_{p}$ has been reported to range from ion-like~\cite{PhysRevMaterials.1.014401,PhysRevMaterials.3.064405} to electron-like ~\cite{schaack1975magnetics,schaack1976observations,schaack1977magnetic,cheng2020large, luo2023large,PhysRevB.109.155426} values. The second term in Eq.~(\ref{Lag}) can be interpreted as the paramagnetic interaction of charged particles $-\mathbf{A}\cdot\mathbf{j}_{p}$, where $\mathbf{A}=\frac{1}{2}\mathbf{B}\times\mathbf{r}$ is the magnetic vector potential and $\mathbf{j}_{p}$ a displacement charge current~\cite{notepara}. 


A ferroelectric with a spontaneous electric polarization along the polar ($z$) axis breaks the horizontal mirror symmetry $\sigma_{h}$ of the $xy$ plane while preserving any vertical mirror symmetry $\sigma_{v}$ through the polar axis. The minimal symmetry-allowed free energy density reads
\begin{align}
f[\mathbf{P}]=&\frac{G}{2}\sum_{i}(\boldsymbol{\nabla}P_{i})^2+\frac{\alpha_{\parallel}}{2}P_{z}^{2}+\frac{\beta_{\parallel}}{4}P_{z}^{4}\nonumber\\&+\frac{\alpha_{\perp}}{2}\left(P_{x}^{2}+P_{y}^{2}\right)+\frac{\beta_{c}}{2}P_{z}^2(P_{x}^2+P_{y}^2),\label{Fperp}
\end{align}
where $G$ is the isotropic Ginzburg parameter accounting for the energy cost of polarization gradients, and the remaining coefficients are Landau parameters. Below the Curie temperature, $\alpha_{\parallel}<0$ and $\beta_{\parallel}>0$, leading to energy minima with spontaneous polarization along the $z$ axis. The last term represents the leading biquadratic coupling between longitudinal and transverse polarization components with respect to the polar axis. This coupling encodes the effect of the ferroelectric ordering on transverse polarization fluctuations and typically appears in ferroelectrics with a coupling strength $\beta_{c}$ comparable to $\beta_{\parallel}$~\cite{haun1987thermodynamic,PhysRevB.74.104104}. For static electric and magnetic fields, $\mathbf{E}=E\mathbf{z}$ and $\mathbf{B}=B\mathbf{z}$, the ground-state polarization $P_{0}\mathbf{z}$ minimizes $f[\mathbf{P}]$ such that $\alpha_{\parallel}P_{0}+\beta_{\parallel}P_{0}^{3}=E$ and $P_{0}=\pm\sqrt{-\alpha_{\parallel}/\beta_{\parallel}}$ when $E=0$. The polarization is stable against transverse fluctuations when $\alpha_{\perp}+\beta_{c}P_{0}^2>0$. For simplicity, we disregarded in Eq.~(\ref{Fperp}) the quartic terms $P_{x}^{4}$ and $P_{y}^{4}$~\cite{Px4}, but retain the essential $P_{z}^4$ contribution. Eq.~(\ref{Fperp}) captures the essential features of ferroelectrics and, when $\alpha_{\parallel}>0$, also describes para- and dielectric systems.


\emph{Second quantization---}The quasiparticle excitations represent small spatio-temporal polarization fluctuations $\mathbf{p}(\mathbf{r},t)=\mathbf{P}(\mathbf{r},t)-P_{0}\mathbf{z}$ on top of the ground-state polarization. The Legendre transform, $\mathcal{H}=\partial_{t}\mathbf{p}\cdot\boldsymbol{\pi} -\mathcal{L}$, yields the Hamiltonian density for $\mathbf{p}(\mathbf{r},t)$, where $\boldsymbol{\pi}\equiv\partial\mathcal{L}/\partial(\partial_{t}\mathbf{p})$ is the canonical momentum conjugate to $\mathbf{p}(\mathbf{r},t)$. To investigate the quasiparticle properties, we quantize the $j$th polarization fluctuation component in terms of boson operators $\hat{a}_{\mathbf{q}j}^{(\dagger)}$. In the Supplemental Material~\cite{SM}, we derive the noninteracting part of the Hamiltonian as
\begin{align}
\hat{H}_{0}=&\hbar\sum_{\mathbf{q}}\big\{\omega_{\mathbf{q}\perp}\big(\hat{a}_{\mathbf{q}x}^{\dagger}\hat{a}_{\mathbf{q}x}+\hat{a}_{\mathbf{q}y}^{\dagger}\hat{a}_{\mathbf{q}y}\big)+\omega_{\mathbf{q}\parallel}\hat{a}_{\mathbf{q}z}^{\dagger}\hat{a}_{\mathbf{q}z}\nonumber\\
&+i\gamma_{p}B\big(\hat{a}_{\mathbf{q}x}^{\dagger}\hat{a}_{\mathbf{q}y}-\hat{a}_{\mathbf{q}y}^{\dagger}\hat{a}_{\mathbf{q}x}\big)\big\}\nonumber\\
=&\hbar\sum_{\mathbf{q}}\big\{\omega_{\mathbf{q}\pm}\hat{a}_{\mathbf{q}\pm}^{\dagger}\hat{a}_{\mathbf{q}\pm} +\omega_{\mathbf{q}\parallel}\hat{a}_{\mathbf{q}z}^{\dagger}\hat{a}_{\mathbf{q}z}\big\}\label{Ham1}
\end{align}
where $\omega_{\mathbf{q}\perp}=\sqrt{v_c^2q^2+\Delta_{\perp}^2+(\gamma_{p}B)^2}$, $\omega_{\mathbf{q}\parallel}=\sqrt{v_{c}^2q^2+\Delta_{\parallel}^2}$, $v_{c}=\sqrt{G/m_{p}}$ is a characteristic velocity, and $\Delta_{\parallel(\perp)}=(m_{p}\chi_{\parallel(\perp)})^{-1/2}$ is the intrinsic frequency gap of the longitudinal (transverse) fluctuation component, with $\chi_{\parallel}=1/(\alpha_{\parallel}+3\beta_{\parallel}P_{0}^{2})$ and $\chi_{\perp}=1/(\alpha_{\perp}+\beta_{c}P_{0}^{2})$ the corresponding static susceptibilities. The external static electric field enters only via the spontaneous polarization $P_{0}(E)$ in the susceptibilities. In the second equality, we transformed to the circularly polarized basis $\hat{a}_{\mathbf{q}\pm}^{\dagger}=(\hat{a}_{\mathbf{q}x}^{\dagger}\pm i\hat{a}_{\mathbf{q}y}^{\dagger})/\sqrt{2}$ of right- ($+$) and left-handed ($-$) transverse modes with eigenfrequencies $\omega_{\mathbf{q}\pm}=\omega_{\mathbf{q}\perp}\mp\gamma_{p}B$. The Zeeman-like frequency splitting represents a~\emph{paramagnetic} contribution arising from the orbital magnetic moments of the circular polarization fluctuations, similar to that of phonons~\cite{PhysRevMaterials.1.014401,PhysRevMaterials.3.064405}, whereas the quadratic dependence of $\omega_{\mathbf{q}\perp}$ on \textit{B} reflects the \emph{diamagnetism} of the polarization dynamics, akin to the orbital diamagnetism of electrons. The anharmonic potentials in Eq.~(\ref{Fperp}) give rise to the interaction \cite{SM}
\begin{align}
\hat{H}_{\text{int}}=&\frac{P_{0}}{\sqrt{V}}\left(\frac{\hbar}{m_{p}}\right)^{3/2}\sum_{\mathbf{k}\mathbf{q}}\hat{A}_{-(\mathbf{k}+\mathbf{q})z}\left\{\beta_{\parallel}\hat{A}_{\mathbf{k}z}\hat{A}_{\mathbf{q}z}\right.\nonumber\\
&\left.+\beta_{c}\left(\hat{A}_{\mathbf{k}x}\hat{A}_{\mathbf{q}x}+\hat{A}_{\mathbf{k}y}\hat{A}_{\mathbf{q}y}\right)\right\}+\mathcal{O}(\hat{A}^{4})\label{Inter}
\end{align}
where $V$ is the system volume and $\hat{A}_{\mathbf{q}j}= (\hat{a}_{\mathbf{q}j}+\hat{a}_{-\mathbf{q}j}^{\dagger})/\sqrt{2\omega_{\mathbf{q}j}}$. We discarded fourth-order contributions, assuming sufficiently small-amplitude fluctuations. The $\beta_{\parallel}$ and $\beta_{c}$ interaction terms originate from the cubic anharmonicities of the polarization fluctuations, $p_{z}^3$ and $p_{z}(p_{x}^2+p_{y}^2)$, respectively, which are odd under the parity operation $p_{z}\rightarrow-p_{z}$. The former turns out to transform transverse circularly polarized modes in Eq.~(\ref{Ham1}) into multiferrons with both intrinsic electric and magnetic dipoles, while the latter leads to conventional ferrons~\cite{PhysRevB.106.L081105, bauer2022magnonics,PhysRevApplied.20.050501, PhysRevB.109.L060301}. These terms vanish in the phase transition to the \emph{paraelectric} phase above the Curie temperature that recovers inversion symmetry. 

\begin{figure} \centering \par \includegraphics[width=5.2 cm]{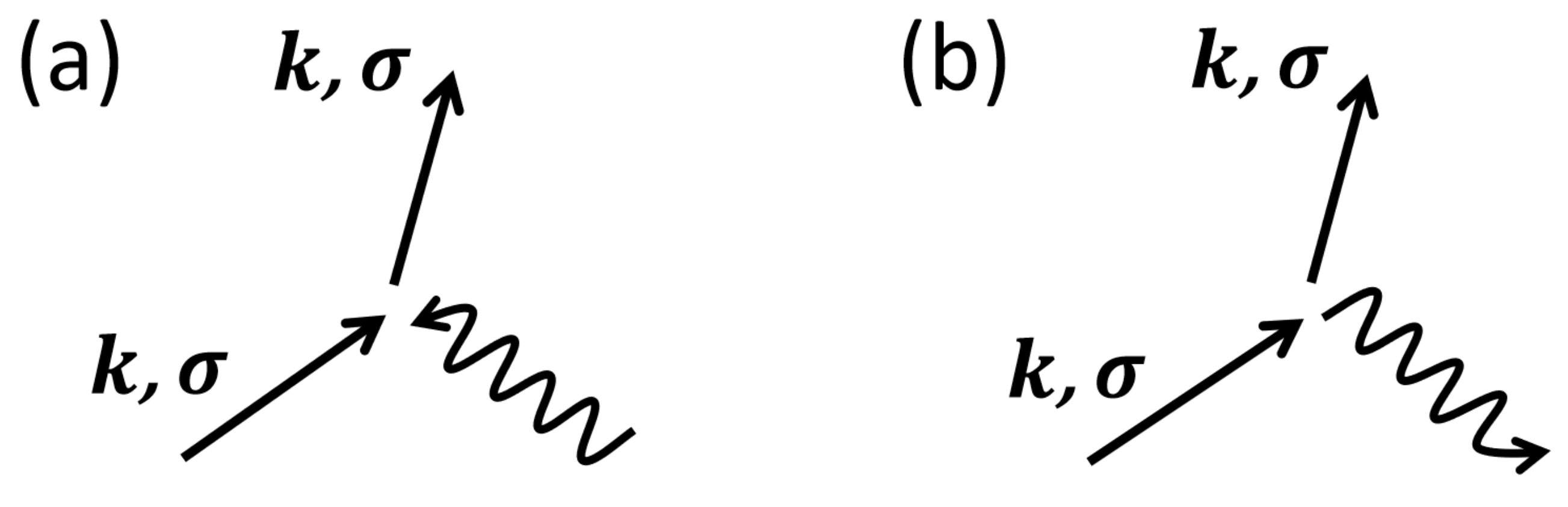} \newline \caption{The parity-odd three-body interactions by which the quasiparticles of mode~$\mathbf{k}\sigma$ acquire an intrinsic electric dipole moment, i.e., by absorbing (a) and emitting (b) a spatially uniform mode (denoted by wave lines) polarized along the ferroelectric order. }\label{Fig-Feyman} 
\end{figure}

\emph{Intrinsic electric and magnetic dipoles---}The axial symmetry around the ferroelectric order allows static dipole moments only along the $z$ axis. The electric dipole of a single-particle excited state $\vert\mathbf{k},\sigma\rangle=\hat{a}_{\mathbf{k}\sigma}^{\dagger}\vert 0\rangle$ modifies the global polarization by
\begin{align}
d_{\mathbf{k}\sigma}=\int d\mathbf{r} \left\{\big\langle \mathbf{k},\sigma\big\vert\hat{p}_{z}(\mathbf{r},t)\big\vert \mathbf{k},\sigma\big\rangle -\big\langle 0\big\vert \hat{p}_{z}(\mathbf{r},t)\big\vert 0\big\rangle \right\} \label{pks}
\end{align}
where $\sigma\in\{\pm,z\}$ labels the quasiparticle branch, and the second term removes the zero-point contribution of the ground state $\vert 0\rangle$. Eq.~(\ref{pks}) involves calculating the expectation value of $\hat{A}_{\mathbf{0}z}(t)$, which can be accessed by solving Eq.~(\ref{Ham1}) and Eq.~(\ref{Inter}) by many-body perturbation theory. For a dilute quasiparticle gas, we may retain only the leading-order term in the interaction, leading to for the circularly (multiferron) and longitudinally (ferron) polarized modes~\cite{SM}
\begin{align}
d_{\mathbf{k}\pm}=\frac{d_{\perp}}{\sqrt{1+k^2l_\perp^2+(B/B_{c})^2}},\,\,\, d_{\mathbf{k}z}=\frac{d_{\parallel}}{\sqrt{1+k^2l_{\parallel}^2}}\label{dipmoment}
\end{align}
with $d_{\perp}=-\hbar\beta_{c}P_{0}\chi_{\parallel}/(m_{p}\Delta_{\perp})$ and $d_{\parallel}=-3\hbar\beta_{\parallel}P_{0}\chi_{\parallel}/(m_{p}\Delta_{\parallel})$. $B_{c}=\Delta_{\perp}/\gamma_{p}$ is a characteristic magnetic field, and $l_{\parallel(\perp)}=\sqrt{G\chi_{\parallel(\perp)}}$ is the polarization coherence length for the longitudinal (transverse) component. The dipole moment of $\mathbf{k}\sigma$-(multi)ferrons arises from the parity-odd three-body interaction of the form $\hat{A}_{\mathbf{0}z}\hat{A}_{\mathbf{k}\sigma}\hat{A}_{-\mathbf{k}\sigma}$ in Eq.~(\ref{Inter}), which contributes through absorption and emission of a uniform $(\mathbf{q}=0)$ longitudinal mode, as schematically illustrated in Fig.~\ref{Fig-Feyman}. $d_{\mathbf{k}z}<0$ indicates that the ferron excitations always reduce the macroscopic polarization (assumed along the positive $z$ direction). In contrast, the sign of the dipole $d_{\mathbf{k}\sigma}$ of multiferrons depends on that of the biquadratic coupling strength $\beta_{c}$. By comparing Eq.~(\ref{dipmoment}) with the quasiparticle frequencies, we find that the Hellmann-Feynman theorem holds~\cite{noteHF}:
\begin{equation}
d_{\mathbf{k}\sigma}=-\frac{\partial\hbar\omega_{\mathbf{k}\sigma}}{\partial E} \label{HF}
\end{equation}
where $\omega_{\mathbf{q}\sigma}$ depends on the electric field through the polarization-dependent susceptibilities. The electric dipole of (multi)ferrons can therefore be acquired directly from the field derivative of their (harmonic) frequencies without the need to explicitly solve the many-body problem. Moreover, a finite intrinsic dipole moment at $E=0$ causes a linear dc Stark effect. Since the (multi)ferrons transport energy in a temperature gradient, they enable electric-field control of the thermal conductivity in ferroelectrics~\cite{PhysRevB.106.L081105,wooten2023electric,zhao2025thermal}, thereby offering opportunities for active thermal management~\cite{wooten2023electric}.

The magnetic dipole, on the other hand, is encoded in the response of the quasiparticle dispersions to an external magnetic field. The Hellmann-Feynman theorem \cite{noteHF} yields the magnetic moments of multiferron modes with opposite handedness as
\begin{align}
\mu_{\mathbf{k}\pm}=&-\frac{\partial \hbar\omega_{\mathbf{k}\pm}}{\partial B}=\pm\mu_{M}-\frac{\mu_{M}(B/B_{c})}{\sqrt{1+k^2l_\perp^2+(B/B_{c})^2}} \label{Mag}
\end{align} 
with $\mu_{M}=\hbar\gamma_{p}$. The first term represents a paramagnetic contribution associated with the intrinsic angular momentum or ``spin" of the transverse polarization fluctuations~\cite{riseborough2010quantum,notespin},
\begin{align}
\hat{s}_{z}= &\frac{1}{2}\int \left(\hat{\mathbf{p}}\times\hat{\boldsymbol{\pi}} - \hat{\boldsymbol{\pi}}\times \hat{\mathbf{p}}\right)_{z} d\mathbf{r}\nonumber\\
=&\hbar\sum_{\mathbf{k}}(\hat{a}_{\mathbf{k}+}^{\dagger}\hat{a}_{\mathbf{k}+}-\hat{a}_{\mathbf{k}-}^{\dagger}\hat{a}_{\mathbf{k}-}), \label{Spin} 
\end{align}
which endows the right- and left-handed multiferrons with angular momenta of $\pm\hbar$ and corresponding magnetic moments of $\pm\mu_{M}$, respectively. Eq.~(\ref{Spin}) is analogous to the spin angular momentum of circularly polarized photons~\cite{andrews2012angular}, but the latter lack a magnetic moment. The field-dependent term in Eq.~(\ref{Mag}) is always negative for $B>0$ and therefore reveals a diamagnetic correction. Beyond the Stark and Zeeman effects, we discuss below other physical consequences of multiferrons in ferroelectrics. 
 
\emph{Second-harmonic generation---}The second-harmonic generation (SHG) refers to the emission of photons at twice the frequency of an incident light and serves as a powerful optical probe of inversion-symmetry breaking in materials~\cite{denev2011probing,matlack2015review}. SHG is typically associated with electronic transitions in noncentrosymmetric compounds at optical (eV) frequencies. Here we show that multiferrons themselves generate a giant second-harmonic response when resonantly driven (at THz frequencies), owing to their intrinsic dipole moments that break the inversion symmetry. Consider a uniform monochromatic electric field $\boldsymbol{\mathcal{E}}\exp(-i\omega t)$. The second-harmonic response $p_{i}(2\omega)=\epsilon_{0}\sum_{jl}\chi_{ijl}^{(2)}(-2\omega;\omega,\omega)\mathcal{E}_{j}\mathcal{E}_{l}$ is governed by the SHG susceptibility tensor $\chi_{ijl}^{(2)}=\chi_{ilj}^{(2)}$, where $\epsilon_{0}$ is the vacuum permittivity and $p_{i}(2\omega)$ is the $2\omega$ Fourier component of the induced polarization. Solving the Landau-Khalatnikov-Tani equation in the absence of a magnetic field~\cite{sivasubramanian2004physical}, $m_{p}\left(\partial_{t}^2\mathbf{p}+\Gamma\partial_{t}\mathbf{p}\right)=-\partial f/\partial\mathbf{p}+\boldsymbol{\mathcal{E}}e^{-i\omega t}$, yields $\chi_{xxz}^{(2)}=2\chi_{zxx}^{(2)}$ and
\begin{subequations}\label{SHG}
 \begin{align}
\chi_{zxx}^{(2)}=&\frac{d_{\perp}}{\epsilon_{0}\hbar\chi_{\perp}\Delta_{\perp}} \frac{\Delta_{\parallel}^2 }{\Delta_{\parallel}^2-\omega^2-i\omega\Gamma}\widetilde{\chi}_{\perp}^2(\omega)\label{xx} , \\ 
\chi_{zzz}^{(2)}=&\frac{d_{\parallel}}{\epsilon_{0}\hbar\chi_{\parallel}\Delta_{\parallel}} \frac{\Delta_{\parallel}^2}{\Delta_{\parallel}^2-\omega^2-i\omega\Gamma}\widetilde{\chi}_{\parallel}^2(\omega) ,\label{zz}
\end{align}       
\end{subequations}
where $\Gamma$ is a phenomenological damping parameter and $\widetilde{\chi}_{\parallel(\perp)}(\omega)=\chi_{\parallel(\perp)}\Delta_{\parallel(\perp)}^{2}/(\Delta_{\parallel(\perp)}^{2}-\omega^{2}-i\omega\Gamma )$ is the linear-response dynamic susceptibility. Axial symmetry imposes $\chi_{zxx}^{(2)}=\chi_{zyy}^{(2)}$, while all other tensor components vanish. Eq.~(\ref{SHG}) shows that the second-harmonic response to a transversely and longitudinally polarized driving field scales with the intrinsic dipole moment of multiferrons and ferrons, and is therefore expected to be absent in conventional phonons without intrinsic (dc) dipole moments.

\emph{Magnetoelectric response---}Magnetoelectric coupling describes a cross response of electric polarization (magnetization) to an applied magnetic (electric) field~\cite{fiebig2005reviva}. Since multiferrons combine intrinsic electric and magnetic dipoles, they respond to both electric and magnetic fields with a finite magnetoelectric coupling coefficient $\alpha_{\text{ME}}=\mu_{0}\partial\langle\delta P\rangle/\partial B$, where $\mu_{0}$ is the vacuum permeability and
\begin{equation}
\langle\delta P\rangle=\int\frac{d\mathbf{k}}{(2\pi)^3}\sum_{\sigma=\pm}\frac{d_{\mathbf{k}\sigma}}{e^{\hbar\omega_{\mathbf{k}\sigma}/k_{B}T}-1}
\end{equation}
is the average change of the macroscopic polarization caused by thermally excited multiferrons at temperature $T$. The magnetoelectric coupling vanishes in the ferroelectric ground state at zero temperature and above the Curie temperature (where $d_{\mathbf{k}\sigma}=0$), but becomes finite in the presence of thermal multiferrons. Since two multiferron branches of opposite handedness are energetically degenerate, $\alpha_{\text{ME}} \sim B$ for small bias magnetic fields.

\emph{Einstein-de Haas effect---}The application of a magnetic field along the ferroelectric order generates a net angular momentum by lifting the degeneracy of two multiferron modes with angular momentum $\pm\hbar$, leading to a mechanical rotation torque on the ferromagnetic sample, i.e., the Einstein-de Haas effect, as observed in magnetic systems~\cite{einstein1915experimental}. This effect is measured by the coefficient $\alpha_{\text{EdH}}=\partial \langle\delta S\rangle/\partial B$, where 
\begin{equation}
\langle\delta S\rangle=\frac{V_{0}}{(2\pi)^{3}}\int d\mathbf{k}\sum_{\sigma=\pm}\frac{\sigma\hbar}{e^{\hbar\omega_{\mathbf{k}\sigma}/k_{B}T}-1} 
\end{equation}
is the induced angular momentum of multiferrons per unit-cell volume $V_{0}$. In contrast to $\alpha_{\text{ME}}$, $\alpha_{\text{EdH}}$ remains finite at zero magnetic field.
\begin{table}
\caption{\label{table1}Calculated gaps ($\Delta$), intrinsic electric and magnetic dipole moments, and room-temperature Einstein-de Haas coefficient ($\alpha_{\text{EdH}}$) of multiferrons and conventional ferrons for BaTiO$_3$ in the absence of external electric and magnetic fields. }
\begin{ruledtabular}
\begin{tabular}{lccccc}
 & $\Delta/2\pi$ (THz) & $d$ (e\AA) & $\mu_M$ ($\mu_{B}$) & $\alpha_{\text{EdH}}$ ($\hbar$/T) \\
\hline
Multiferrons & 0.74 & $-0.34$ & $6.6\times 10^{-3}$ & $1.2\times 10^{-5}$  \\
Ferrons      & 3.1  & $-1.44$\footnotemark[1] & \ding{55} &  \ding{55}   \\
\end{tabular}
\end{ruledtabular}
\footnotetext[1]{approximately half of that in Ref.~\cite{PhysRevB.106.L081105}, in which a higher-order anharmonic Landau potential was employed.}
\end{table}
\begin{figure}
    \centering
    \includegraphics[width=8.2 cm]{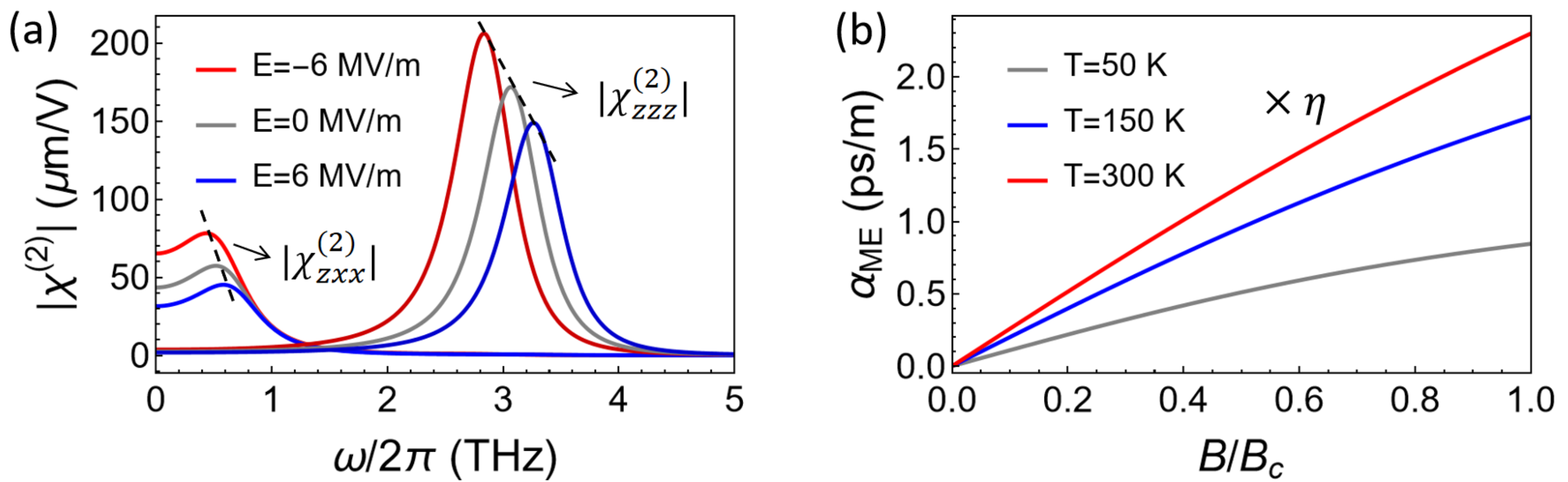}
    \caption{ (a) Electric-field-tunable second-harmonic generation by multiferrons ($\vert\chi_{zxx}^{(2)}\vert$) and ferrons ($\vert\chi_{zzz}^{(2)}\vert$). (b) Multiferron-induced magnetoelectric coupling coefficient multiplied by $\eta$ as a function of the bias magnetic field ($B$). Here $\eta=\gamma_{e}/\gamma_{p}$ denotes the ratio of the gyromagnetic ratio of electrons to that of the electric polarization. }
    \label{Fig-ME}
\end{figure}

Finally, we estimate the magnitude of the predicted effects for the common ferroelectric BaTiO$_3$ with model parameters~\cite{noteparameters}: $\alpha_{\parallel}=-2.58\times 10^8\,$Vm/C, $\alpha_{\perp}=-1.46\times 10^7\,$Vm/C, $\beta_{\parallel}=3.82\times 10^9\,$Vm$^5$/C$^3$, $\beta_{c}=6.46\times10^{8}\,$Vm$^5$/C$^3$, $G=1.83\times10^{-10}\,$Vm$^3$/C, $m_{p}=1.35\times10^{-18}\,$Vms$^2$/C~\cite{PhysRevB.106.L081105}, and $\Gamma/2\pi=0.76\,$THz~\cite{Gamma}. Fig.~\ref{Fig-ME}(a) presents the absolute values of the calculated SHG susceptibilities $\chi_{zxx}^{(2)}$ and $\chi_{zzz}^{(2)}$ that exhibit electric-field-tunable resonant enhancements near the multiferron and ferron frequencies, respectively. Moreover, $\vert\chi_{zzz}^{(2)}\vert$ exceeds $\vert\chi_{zxx}^{(2)}\vert$ due to a larger dipole moment of ferrons [see Table~\ref{table1}]. These values are approximately \emph{five} to \emph{six} orders of magnitude larger than those reported in 2D nonlinear materials with record SHG in the optical (eV-scale) regime~\cite{abdelwahab2022giant,li2025giant}, highlighting the potential of (multi)ferrons for THz nonlinear optical applications~\cite{zhang2025electric}. Fig.~\ref{Fig-ME}(b) shows that the multiferron-induced $\alpha_{\text{ME}}$ increases approximately linearly with the bias magnetic field $B$. When $B\sim B_{c}$ it approaches values about $\gamma_{p}/\gamma_{e}$ times that of the prototypical magnetoelectric antiferromagnet Cr$_2$O$_3$~($\sim$ps/m)~\cite{astrov1961magnetoelectric,PhysRevLett.7.310,PhysRevLett.6.607}, where $\gamma_{e}=-1.76\times10^{11}\,$T$^{-1}$s$^{-1}$ is the gyromagnetic ratio of electrons. We estimate $\gamma_{p}$ for BaTiO$_3$ by including both ionic and electronic contributions, $\gamma_{p}=\gamma_{\text{ion}}+\gamma_{\text{ele}}$. The ionic part can be estimated as that of the circular motion of Ti ions in the oxygen octahedron, i.e., $\gamma_{\text{ion}}\simeq Z_{\text{Ti}}^{\ast}/(2M_{\text{Ti}})$, and using the Born effective charge $Z_{\text{Ti}}^{\ast}=+7.25e$~\cite{PhysRevB.51.6765} yields $\gamma_{\text{ion}}\simeq7.3\times 10^{6}\,$T$^{-1}$s$^{-1}$. $\gamma_{\text{ele}}$ is caused by electron-phonon coupling and can be evaluated within second-order perturbation theory as $\gamma_{\text{ele}}\simeq(g_{J}\mu_{0}\mu_{B})^{-1}V_{0}\Delta\vert g_{\text{e-p}}\vert^2/(E_{g}^2-\hbar^2\Delta^2)$~\cite{notere}, where $g_{J}\sim1$, $g_{\text{e-p}}$, and $E_{g}$ are the electronic Land\'e factor, electron-coupling strength, and electronic band gap, respectively. Using $g_{J}\sim1$, $V_{0}=66\,$\AA$^3$, $E_g=3.2\,$eV, and $g_{\text{e-p}}=15\,$meV~\cite{noteg}, we obtain $\gamma_{\text{ele}}=5.77\times 10^8\,$T$^{-1}$s$^{-1}$, leading to $\gamma_{p}= 5.84\times 10^8\, \mathrm{T}^{-1}\mathrm{s}^{-1}$, which is about one order of magnitude smaller than that of soft chiral phonons reported in PbTe~\cite{PhysRevLett.128.075901}. Table~\ref{table1} compares the calculated properties of multiferrons and ferrons. The small value of $\gamma_{p}$ yields $\eta\simeq-3.3\times 10^{-3}$, $B_{c}=7.94\times 10^{3}\,$T, $\mu_{M}=6.6\times 10^{-3}\,\mu_{B}$, and $\alpha_{\text{EdH}}\simeq 1.2\times10^{-5}\,\hbar$/T at room temperature. For $B=10\,$T, this implies $\alpha_{\text{ME}}\sim -10^{-5}\,$ps/V, a Zeeman splitting $\gamma_{p}B\simeq 0.9\,$GHz, and $\langle\delta S\rangle\sim10^{-4}\,\hbar$ per unit cell, which are relatively small but should not be unobservable. For instance, $\langle\delta S\rangle\sim 10^{-4}\,\hbar$ corresponds to a torque density $\langle\delta S\rangle/(\tau_{c}V_{0})\sim 0.1\,$N/m$^2$ for a characteristic magnetic-field ramp time $\tau_{c}\sim1\,$ns, comparable to the torque recently measured in tellurium driven by chiral phonons~\cite{PhysRevLett.121.175301,zhang2025measurement}. In addition, while our estimate focuses on the displacive ferroelectric BaTiO$_3$, we note that in other materials $\gamma_{p}$ can be significantly enhanced up to the order of $\gamma_{e}$ by a large electronic contribution, e.g., in rare-earth halides with a small relevant $E_{g}$~\cite{schaack1975magnetics,schaack1976observations,schaack1977magnetic,luo2023large, PhysRevLett.133.266702}. In contrast to their small magnetic moment and Zeeman effect, multiferrons possess a sizable intrinsic dipole moment of $-0.34\,$e\AA{}, leading to a blueshift of $8.2\,$GHz in their frequencies under a moderate electric field of $10\,$kV/cm applied along the spontaneous polarization.

\emph{Conclusions---}We predict a new class of quasiparticle excitations in ferroelectrics. Because they carry \emph{static} electric as well as magnetic dipoles we call them ``multiferrons". The magnetic moment is associated with circularly polarized transverse fluctuations of the ferroelectric order. In contrast to conventional chiral phonons, they generate an electric-field-controllable second-harmonic response that is resonantly enhanced in the THz-frequency regime to exceptionally high values. Moreover, they cause a finite cross-magnetoelectric coupling and the Einstein-de Haas effect in ferroelectrics.  
The present model may be extended by incorporating piezoelectric coupling to lattice strains, which could hybridize multiferrons with (acoustic) phonons to form electro- and magneto-active hybrid quasiparticles~\cite{PTunpublished}. While we focus on ferroelectrics, the multiferron concept may apply more broadly to other systems, e.g., magnetic polar materials, in which large gyromagnetic ratios and related effects of multiferrons are expected. Our findings open new opportunities for nonlinear THz photonics and multifunctional magnetoelectric devices based on multiferrons.

\emph{Note added---}While preparing this manuscript, we became aware of a closely related study~\cite{noteadded}.

\emph{Acknowledge---}P.T. and G.B. acknowledge financial support by JSPS KAKENHI Grants Nos. 22H04965 and 24H02231.

\bibliography{reference}

\begin{thebibliography}{99}
\bibitem [{\citenamefont {{The paramagnetic interaction
  $-\mathbf{A}\cdot\mathbf{j}_{p}$ of charged particles reduces to the form
  $\mathbf{B}\cdot(\mathbf{P}\times\partial_{t}\mathbf{P})$ upon substituting
  $\mathbf{j}_{p}\rightarrow\partial_{t}\mathbf{P}$ and
  $q\mathbf{r}\rightarrow\mathbf{P}$}}()}]{notepara}%
  \BibitemOpen
  \bibfield  {author} {\bibinfo {author} {\bibnamefont {{The paramagnetic
  interaction $-\mathbf{A}\cdot\mathbf{j}_{p}$ of charged particles reduces to
  the form $\mathbf{B}\cdot(\mathbf{P}\times\partial_{t}\mathbf{P})$ upon
  substituting $\mathbf{j}_{p}\rightarrow\partial_{t}\mathbf{P}$ and
  $q\mathbf{r}\rightarrow\mathbf{P}$}}},\ }\href@noop {} {}\bibinfo {note}
  {{where $q$ is the charge of particles.}}\BibitemShut {Stop}%
\bibitem [{Px4()}]{Px4}%
  \BibitemOpen
  \href@noop {} {}\bibinfo {note} {{Quartic terms $P_{x}^4$ and $P_{y}^4$
  introduce a parity-even anharmonicity for transverse polarization
  fluctuations, which modifies only the dispersion relation but, by symmetry,
  cannot generate static electric dipoles of quasiparticles.}}\BibitemShut
  {Stop}%
\bibitem [{\citenamefont {{According to the Hellmann-Feynman theorem
  $\langle\phi\vert \hat{\mathbf{P}}\vert\phi\rangle=-\partial\langle\phi\vert
  \hat{H}\vert\phi\rangle/\partial\mathbf{E}$ }}()}]{noteHF}%
  \BibitemOpen
  \bibfield  {author} {\bibinfo {author} {\bibnamefont {{According to the
  Hellmann-Feynman theorem $\langle\phi\vert
  \hat{\mathbf{P}}\vert\phi\rangle=-\partial\langle\phi\vert
  \hat{H}\vert\phi\rangle/\partial\mathbf{E}$ }}},\ }\href@noop {} {}\bibinfo
  {note} {{where $\hat{\mathbf{P}}=-\partial\hat{H}/\partial\mathbf{E}$ is the
  electric polarization operator, $\hat{H}$ is the Hamiltonian under an
  external electric field $\mathbf{E}$ with eigenstates $\vert\phi\rangle$. For
  the states $\vert\mathbf{k},\sigma\rangle$ and $\vert 0\rangle$,
  respectively, we obtain
  $d_{\mathbf{k}\sigma}=-\partial\hbar\omega_{\mathbf{k}\sigma}/\partial E$,
  where
  $\hbar\omega_{\mathbf{k}\sigma}=\langle\mathbf{k},\sigma\vert\hat{H}\vert\mathbf{k},\sigma\rangle-\langle
  0\vert\hat{H}\vert 0\rangle$ is the excitation energy of a single
  quasiparticle of mode $\mathbf{k}\sigma$. Analogously, the magnetic moment
  carried by a multiferron equals the negative of the derivative of its
  frequency with respect to an applied magnetic field}}\BibitemShut {NoStop}%
\bibitem [{not()}]{notespin}%
  \BibitemOpen
  \href@noop {} {}\bibinfo {note} {{According to the canonical commutator
  $[\hat{p}_{i}(\mathbf{r},t), \hat{\pi}_{j}(\mathbf{r}^{\prime}, t)]=i\hbar
  \delta_{ij}\delta(\mathbf{r}-\mathbf{r}^{\prime})$, the spin of the
  polarization fluctuation field given by
  $\hat{\mathbf{s}}=\frac{1}{2}\int(\hat{\mathbf{p}}\times\hat{\boldsymbol{\pi}}-\hat{\boldsymbol{\pi}}\times\hat{\mathbf{p}})d\mathbf{r}$
  indeed satisfies the standard commutation relation of angular momentum
  $[\hat{s}_{i}, \hat{s}_{j}]=i\hbar\sum_{l}\epsilon_{ijl}\hat{s}_{l}$, where
  $\epsilon_{ijl}$ is the three-dimensional antisymmetric tensor.}}\BibitemShut
  {Stop}%
\bibitem [{\citenamefont {{The LGD parameters for BaTiO$_3$ are deduced from
  Ref.~\cite{PhysRevB.74.104104} as
  $\alpha_{\parallel}=-(\alpha_{1}+6\alpha_{11}P_{0}^{2}+15\alpha_{111}P_{0}^4)$}}()}]{noteparameters}%
  \BibitemOpen
  \bibfield  {author} {\bibinfo {author} {\bibnamefont {{The LGD parameters for
  BaTiO$_3$ are deduced from Ref.~\cite{PhysRevB.74.104104} as
  $\alpha_{\parallel}=-(\alpha_{1}+6\alpha_{11}P_{0}^{2}+15\alpha_{111}P_{0}^4)$}}},\
  }\href@noop {} {}\bibinfo {note}
  {{$\beta_{\parallel}=-\alpha_{\parallel}/P_{0}^2$,
  $\alpha_{\perp}=2\alpha_{1}+2\alpha_{112}P_{0}^4$, $\beta_{c}=2\alpha_{12}$,
  and $G=(G_{11}+2G_{44})/3$ with $P_{0}=26\,\mu$m/cm$^{2}$}}\BibitemShut
  {NoStop}%
\bibitem [{\citenamefont {{The electron-phonon coupling includes an electronic
  energy splitting of $\delta E=2\hbar\Delta\vert
  g_{\text{e-p}}\vert^2/(E_{g}^2-\hbar^2\Delta^2)$ in the $p$-orbital valence
  band \cite{PhysRevLett.133.266702}}}()}]{notere}%
  \BibitemOpen
  \bibfield  {author} {\bibinfo {author} {\bibnamefont {{The electron-phonon
  coupling includes an electronic energy splitting of $\delta
  E=2\hbar\Delta\vert g_{\text{e-p}}\vert^2/(E_{g}^2-\hbar^2\Delta^2)$ in the
  $p$-orbital valence band \cite{PhysRevLett.133.266702}}}},\ }\href@noop {}
  {}\bibinfo {note} {{which corresponds to the Zeeman splitting under an
  effective magnetic field generated by a phonon-induced magnetization of the
  order $M\sim \delta E/(2\mu_{0}g_{J}\mu_{B})$. Our estimate of
  $\gamma_{\text{ele}}$ differs from the that in Ref.~\cite{7lm1-wm3y} by a
  factor of the material's magnetic susceptibility}}\BibitemShut {NoStop}%
\bibitem [{\citenamefont {{The electron-phonon coupling in BaTiO$_3$ is
  measured by the dimensionless coupling parameter
  $g_{\text{e-p}}/\hbar\Omega\sim
  3.8-4.6$~\cite{PhysRevB.94.195131}}}()}]{noteg}%
  \BibitemOpen
  \bibfield  {author} {\bibinfo {author} {\bibnamefont {{The electron-phonon
  coupling in BaTiO$_3$ is measured by the dimensionless coupling parameter
  $g_{\text{e-p}}/\hbar\Omega\sim 3.8-4.6$~\cite{PhysRevB.94.195131}}}},\
  }\href@noop {} {}\bibinfo {note} {{where $\Omega$ is the relevant phonon
  frequency. For a multiferron frequency $\Delta/2\pi=0.74\,$THz,
  $g_{\text{e-p}}\sim 15\,$meV, which is around twice that for the soft phonon
  mode reported in SrTiO$_3$~\cite{PhysRevLett.133.266702}}}\BibitemShut
  {NoStop}%
\end{thebibliography}
\end{document}


\title{Supplemental Material for ``Multiferroic-like Quasiparticles in Ferroelectrics"}

\author{Ping Tang}
\email{tang.ping.a2@tohoku.ac.jp}	
\affiliation{WPI-AIMR, Tohoku University, 2-1-1 Katahira, Sendai 980-8577, Japan}
\author{Gerrit E. W. Bauer}

\affiliation{WPI-AIMR, Tohoku University, 2-1-1 Katahira, Sendai 980-8577, Japan}
\affiliation{Institute for Materials Research and CSIS, Tohoku University,
2-1-1 Katahira, Sendai 980-8577, Japan} 
\affiliation{Kavli Institute for Theoretical Sciences, University of the Chinese Academy of Sciences,
Beijing 10090, China}
\date{\today}

\maketitle
In the Supplemental Material, we present the derivation of the Hamiltonian for the polarization fluctuations as well as calculate the electric dipole moments of multiferrons based on many-body perturbation theory.  

\section{The Hamiltonian of interacting quasiparticles}
We introduce small spatio-temporal polarization fluctuations $\mathbf{p}(\mathbf{r},t)$ on top of the ground-state polarization. Substituting $\mathbf{P}(\mathbf{r},t)=P_{0}\mathbf{z}+\mathbf{p}(\mathbf{r},t)$ into the Lagrangian density Eq.~(1) in the main text and using the Legendre transform, $\mathcal{H}=\int\left(\partial_{t}\mathbf{p}\cdot\boldsymbol{\pi} -\mathcal{L}\right)d\mathbf{r}$, leads to the Hamiltonian for the polarization fluctuations:
\begin{align}
H=\int\mathcal{H}\, d\mathbf{r}=\int d\mathbf{r}\bigg\{\frac{1}{2 m_{p}}\big[\boldsymbol{\pi}-\Lambda_{p}\big(\mathbf{B}\times\mathbf{p}\big)\big]^{2} +\frac{G}{2}\sum_{i}(\boldsymbol{\nabla}p_{i})^2+\frac{p_{z}^2}{2\chi_{\parallel}}+\frac{(p_{x}^{2}+p_{y}^{2})}{2\chi_{\perp}}+\delta f [\mathbf{p}]\bigg\},\label{Ham}
\end{align}%
where $\boldsymbol{\pi}\equiv\partial\mathcal{L}/\partial(\partial_{t}\mathbf{p})$ is the canonical momentum conjugate to $\mathbf{p}$, $\chi_{\parallel}=1/(\alpha_{\parallel}+3\beta_{\parallel}P_{0}^{2})$ and $\chi_{\perp}=1/(\alpha_{\perp}+\beta_{c}P_{0}^{2})$ are longitudinal and transverse (static) electric susceptibilities, respectively, and the last term
\begin{equation}
\delta f=\beta_{\parallel}\left(P_{0} +\frac{p_{z}}{4}\right) p_{z}^{3}+\beta_{c}\left(P_{0}+\frac{p_{z}}{2}\right)p_{z}\left(p_{x}^{2}+p_{y}^{2}\right) \label{def}
\end{equation}
is the anharmonic correction. The linear terms in Eq.~(\ref{Ham}) vanish and the external static electric field affects \emph{only} the spontaneous polarization $P_{0}(E)$. The harmonic (quadratic) terms in Eq.~(\ref{Ham}) govern the eigenmodes of the polarization fluctuations while $\delta f$ represents their interactions that include both odd and even terms under the parity operation $p_{z}\rightarrow-p_{z}$. To obtain the quasiparticle Hamiltonian, we quantize the polarization fluctuations as
\begin{align}
\hat{p}_{j}(\mathbf{r},t)=&\sqrt{\frac{\hbar}{2m_{p}V}}\sum_{\mathbf{q}} \frac{\hat{a}_{\mathbf{q}j}}{\sqrt{\omega_{\mathbf{q}j}}}e^{i\mathbf{q}\cdot\mathbf{r}}+\text{H.c.} \label{p}\\
\hat{\pi}_{j}(\mathbf{r},t)=&-i\sqrt{\frac{\hbar m_{p}}{2V}}\sum_{\mathbf{q}}\sqrt{\omega_{\mathbf{q}j}}\hat{a}_{\mathbf{q}j}e^{i\mathbf{q}\cdot\mathbf{r}}+\text{H.c.}\label{pi}  \end{align}
where $V$ is the system volume and $\hat{a}_{\mathbf{q}j}^{(\dagger)}$ annihilates (creates) a bosonic quasiparticle for the $j$th fluctuation component with wave vector $\mathbf{q}$ and frequency $\omega_{\mathbf{q}j}$. The boson commutation relations $[\hat{a}_{\mathbf{q}j},\hat{a}_{\mathbf{q}^{\prime}j^{\prime}}^{\dagger}]=\delta_{\mathbf{q}\mathbf{q}^{\prime}}\delta_{jj^{\prime}}$ and $[\hat{a}_{\mathbf{q}j},\hat{a}_{\mathbf{q}^{\prime}j^{\prime}}]=0$ ensure the canonical commutation $[\hat{p}_{j}(\mathbf{r},t),\hat{\pi}_{j^{\prime}}(\mathbf{r}^{\prime},t)]=i\hbar\delta_{jj^{\prime}}\delta(\mathbf{r}-\mathbf{r}^{\prime})$. Substituting Eq.~(\ref{p}) and Eq.~(\ref{pi}) into Eq.~(\ref{Ham}) leads to the non-interacting (harmonic) part of the Hamiltonian as
\begin{align}
\hat{H}_{0}=&\hbar\sum_{\mathbf{q}}\big\{\omega_{\mathbf{q}\perp}\big(\hat{a}_{\mathbf{q}x}^{\dagger}\hat{a}_{\mathbf{q}x}+\hat{a}_{\mathbf{q}y}^{\dagger}\hat{a}_{\mathbf{q}y}\big)+\omega_{\mathbf{q}\parallel}\hat{a}_{\mathbf{q}z}^{\dagger}\hat{a}_{\mathbf{q}z}+i\gamma_{p}B\big(\hat{a}_{\mathbf{q}x}^{\dagger}\hat{a}_{\mathbf{q}y}-\hat{a}_{\mathbf{q}y}^{\dagger}\hat{a}_{\mathbf{q}x}\big)\big\}\label{Ham1}
\end{align}
where $\omega_{\mathbf{q}\perp}=\sqrt{v_c^2q^2+\Delta_{\perp}^2+(\gamma_{p}B)^2}$, $\omega_{\mathbf{q}\parallel}=\sqrt{v_{c}^2q^2+\Delta_{\parallel}^2}$, $v_{c}=\sqrt{G/m_{p}}$ is a characteristic velocity, and $\Delta_{\parallel(\perp)}=(m_{p}\chi_{\parallel(\perp)})^{-1/2}$ is the intrinsic frequency gap of the $j$th fluctuation component. The anharmonic correction Eq.~(\ref{def}) gives the interaction
\begin{align}
\hat{H}_{\text{int}}=&\frac{P_{0}}{\sqrt{V}}\left(\frac{\hbar}{m_{p}}\right)^{3/2}\sum_{\mathbf{k}\mathbf{q}}\hat{A}_{-(\mathbf{k}+\mathbf{q})z}\left\{\beta_{\parallel}\hat{A}_{\mathbf{k}z}\hat{A}_{\mathbf{q}z}+\beta_{c}\left(\hat{A}_{\mathbf{k}x}\hat{A}_{\mathbf{q}x}+\hat{A}_{\mathbf{k}y}\hat{A}_{\mathbf{q}y}\right)\right\}+\mathcal{O}(\hat{A}^{4})\label{Inter}
\end{align}
with $\hat{A}_{\mathbf{q}j}= (\hat{a}_{\mathbf{q}j}+\hat{a}_{-\mathbf{q}j}^{\dagger})/\sqrt{2\omega_{\mathbf{q}j}}$. Here we discarded fourth-order contributions, assuming sufficiently small-amplitude fluctuations. 

\section{Electric dipole moment of quasiparticles}
The electric dipole of interacting quasiparticles can be accessed by solving Eq.~(\ref{Ham1}) and Eq.~(\ref{Inter}) based on the standard many-body perturbation theory. Because of the rotation symmetry around the ferroelectric order, these quasiparticles can only carry nonzero \text{static} electric dipole moments along the ferroelectric ($z$) axis, which is associated with a uniform shift of the equilibrium polarization and given by the expectation value of $\int\mathbf{z}\cdot\hat{\mathbf{p}}(\mathbf{r},t)d\mathbf{r}$ (i.e., $\hat{A}_{\mathbf{0}z}$). We calculate the dipole expectation value by treating the interaction as a small perturbation. Following the conventional procedure, we assume that the interaction is adiabatically switched on from an infinite past $t_{0}=-\infty$: $\hat{H}=\hat{H}_{0}+e^{-0^{+}\vert t\vert}\hat{H}_{\text{int}}$. The expectation value of $\hat{A}_{\mathbf{0}z}$ reads
\begin{align}
\big{\langle}\mathbf{k},\sigma\big{\vert} \hat{A}_{\mathbf{0}z}(t)\big{\vert} \mathbf{k},\sigma\big{\rangle}={}_0\big{\langle}\mathbf{k},\sigma\vert \hat{S}(-\infty,t)\hat{\mathcal{A}}_{\mathbf{0}z}(t) \hat{S}(t,-\infty)\big{\vert}\mathbf{k},\sigma\big{\rangle}_{0}\label{dip2}
\end{align}
where $\hat{S}(t,t^{\prime})=\hat{T}\exp[-\frac{i}{\hbar}\int_{t^{\prime}}^{t} \hat{H}_{\text{int}}(t^{\prime})dt^{\prime}]$ is the evolution operator. $\hat{A}_{\mathbf{k}\sigma}(t)$ is in the Heisenberg picture, while $\hat{\mathcal{A}}_{\mathbf{k}\sigma}(t)$ is in the interaction picture. They are connected by $\hat{A}_{\mathbf{0}z}(t)=\hat{S}(-\infty,t)\hat{\mathcal{A}}_{\mathbf{0}z}(t)\hat{S}(t,-\infty)$. $\vert\mathbf{k},\sigma\rangle_{0}$ corresponds to an unperturbed (non-interacting) single-particle state with ${}_0\langle\mathbf{k},\sigma\vert \hat{\mathcal{A}}_{\mathbf{k}\sigma}(t)\vert\mathbf{k},\sigma\rangle_{0}=0$. For a dilute quasiparticle gas, we may treat the interaction perturbatively. To leading order, we find
\begin{align}
\langle\mathbf{k},&\sigma\vert\hat{A}_{\mathbf{0}z}(t)\vert \mathbf{k},\sigma\rangle\simeq
\frac{P_{0}}{\sqrt{V}}\frac{\hbar^{3/2}}{m_{p}^{3/2}}\sum_{\mathbf{k}^{\prime}}\int_{-\infty}^{t} dt^{\prime}\mathcal{D}_{\mathbf{0}z}(t-t^{\prime}){}_0\langle \mathbf{k},\sigma\vert\big\{3\beta_{\parallel}\hat{\mathcal{A}}_{\mathbf{k}^{\prime}z}(t^{\prime})\hat{\mathcal{A}}_{-\mathbf{k}^{\prime}z}(t^{\prime})+\beta_{c}[\hat{\mathcal{A}}_{\mathbf{k}^{\prime}x}(t^{\prime})\hat{\mathcal{A}}_{-\mathbf{k}^{\prime}x}(t^{\prime})\nonumber\\& +\hat{\mathcal{A}}_{\mathbf{k}^{\prime}y}(t^{\prime})\hat{\mathcal{A}}_{-\mathbf{k}^{\prime}y}(t^{\prime})]\big\}\vert\mathbf{k},\sigma\rangle_{0} +\mathcal{O}(\hat{\mathcal{A}}^{5})\label{Expan}
\end{align}
where $\mathcal{D}_{\mathbf{0}z}(t-t^{\prime})\equiv(-i/\hbar)\Theta(t-t^{\prime})[\hat{\mathcal{A}}_{\mathbf{0}z}(t), \hat{\mathcal{A}}_{\mathbf{0}z}^{\dagger}(t^{\prime})]=(-1/\hbar)\Theta(t-t^{\prime}) \sin[\omega_{\mathbf{0}z}(t-t^{\prime})]/ \omega_{\mathbf{0}z}$ is the propagator of noninteracting longitudinal modes at $\mathbf{q}=0$. A similar expression holds for the expectation value of $\hat{A}_{\mathbf{0}z}(t)$ over the ground state \(\big{\langle} 0\big{\vert} \hat{A}_{\mathbf{q}z}(t)\big{\vert} 0\big{\rangle}\). In the classical regime, the dipole expectation value Eq.~(\ref{Expan}) of interacting quasiparticles in terms of averages over non-interacting states corresponds to the expansion of the polarization dynamics into harmonic oscillators~\cite{PhysRevB.106.L081105}. After some tedious but straightforward algebra, we find the electric dipoles for transverse and longitudinal modes as
\begin{align}
d_{\mathbf{k}\pm}=-\frac{\beta_{c}\hbar P_{0}}{m_{p}^{2}\omega_{\mathbf{0}z}^{2}}\frac{1}{\omega_{\mathbf{k}\perp}},\,\,\,
d_{\mathbf{k}z}=-\frac{3\beta_{\parallel}\hbar P_{0}}{m_{p}^{2}\omega_{\mathbf{0}z}^{2}}\frac{1}{\omega_{\mathbf{k}z}}
\end{align}
which reduce to those given in the main text upon substituting the explicit expressions of quasiparticle frequencies.

\bibliography{reference}